\documentstyle[11pt]{article}
\begin{document}   
\begin{center}{ \bf  About integrated invariance, arising at resonance 
of oscillations in some types non-conservative mechanical systems}
\end{center}
\noindent
\begin{center}{ \bf A.N. Skripka }
\end{center}
\begin{center}{ \bf KCK Soft, Kiev, Ukraine }
\end{center}
\begin{center}{ skripka@ukrpost.net }
\end{center}

{ \it For system of two ordinary differential equations of the second order 
representing autonomous non-conservative holonomic mechanical system, 
in case of dynamics such as one-frequency periodical oscillations, 
is found integrated invariant of movement.}

\vspace{3mm}

Consider a system of two ordinary differential equations of the second order:
$$
\frac{d^2x}{dt^2}=I(x,y),\qquad \frac{d^2y}{dt^2}=J(x,y).\eqno(1)
$$
It can be interpreted as movement of a particle of unit weight
in two-dimensional space $ (x,y) $ under action of an any force field
$$  \vec F(x,y)=(I(x,y),J(x,y)). $$
Using the Helmholz theorem [1,p177] any vector field is possible to present 
as the sums of potential and vortical fields
$$ \vec F(x,y)=\vec F_1(x,y)+\vec F_2(x,y).
$$
The potential field represents a gradient some scalar field
$ U=U(x,y): $
$$
\vec F_1=-grad\quad U=(-\frac{\partial U}{\partial x},
-\frac{\partial U}{\partial y}).\eqno(2)
$$
For a two-dimensional vortical field
$ \vec F_2=(F_{2x}(x,y),F_{2y}(x,y)) $ 
the expression is fair:
$$
 div \vec F_2=\frac{\partial F_{2x}}{\partial x}+ 
         \frac{\partial F_{2y}}{\partial y}, 
$$
through which this field is possible to define through vortical potential
$  \psi(x,y): $
$$
\vec F_2=(\frac{\partial\psi}{\partial y},
-\frac{\partial\psi}{\partial x}).\eqno(3)
$$
Using expressions (2) and (3) field 
$ \vec F $ is possible to present as follows:
$$
\vec F(x,y)=(-\frac{\partial U}{\partial x}+
\frac{\partial\psi}{\partial y},
-\frac{\partial U}{\partial y}-\frac{\partial\psi}{\partial x}).
$$
Taking into account above mentioned ratio system (1)
is possible to write as:
$$
\frac{d^2x}{dt^2}=-\frac{\partial U}{\partial x}+
\frac{\partial\psi}{\partial y},\qquad
\frac{d^2y}{dt^2}=-\frac{\partial U}{\partial y}-
\frac{\partial\psi}{\partial x}.\eqno(4)
$$
The system (4) represents autonomous non-conservative
holonomic mechanical system.                         

Let enter function
$ H(x,y,p,r)=0.5(p^2+r^2)+U(x,y), p=dx/dt, r=dy/dt, $ 
representing complete mechanical energy of system (4).
Using this function, system (4) is possible to present as:
$$
\frac{dx}{dt}=\frac{\partial H}{\partial p},\qquad
\frac{dp}{dt}=-\frac{\partial H}{\partial x}+
\frac{\partial\psi}{\partial y},\qquad
$$
$$
\frac{dy}{dt}=\frac{\partial H}{\partial r},\qquad
\frac{dr}{dt}=-\frac{\partial H}{\partial y}-
\frac{\partial\psi}{\partial x}.\eqno(5)
$$
Multiplying the first equation (5) on $ dp/dt $,
second equation on $ -dx/dt $, third on
$ dr/dt $, fourth on $ -dy/dt $ and
having combined together four received equations, we shall receive:
$$
\frac{dH}{dt}=
\frac{\partial\psi}{\partial y}\frac{dx}{dt}-
\frac{\partial\psi}{\partial x}\frac{dy}{dt}.\eqno(6)
$$
The expression in the right part (6) is a capacity of non-potential forces,
working in system (4).

By the most common case of regular oscillatory dynamics of system
(4) are two-frequency quasi-periodical oscillations, when          
the trajectories everywhere densely fill on a surface 2-torus in phase space 
$ (x, dx/dt, y, dy/dt). $ 
Thus of frequency of oscillations $ T_1 $ and $ T_2 $
are incommensurable, that is them is impossible to present as
$ T_1/T_2=m/n, $ where $ m, n $ - natural numbers. However at
the certain values of parameters and entry conditions 
the frequencies of oscillations will be corresponds as natural numbers and 
then the trajectories of system (4) will represent 
the closed curve on a surface 2-torus. It is a mode 
occurrence of a resonance, when the system makes one-frequency
oscillations with the period $ T=nT_1=mT_2. $ 
Further given case will be is considered more in detail.

Integrate equation (6) on time in interval from 0 up to $ T: $
$$
\int\limits_0^T (\frac{dH}{dt}+\frac{\partial\psi}{\partial x}
\frac{dy}{dt}-\frac{\partial\psi}{\partial y}\frac{dx}{dt})dt=
H(t=T)-H(t=0)+\int\limits_0^T(\frac{dH}{dt}+\frac{\partial\psi}{\partial x}
\frac{dy}{dt}-\frac{\partial\psi}{\partial y}\frac{dx}{dt})dt=0.
$$
As the considered oscillations are one-frequency with the period
$ T, $ that $ H (t=0)=H (t=T) $ and last expression will accept a kind:
$$
\int\limits_0^T(\frac{\partial\psi}{\partial x}
\frac{dy}{dt}-\frac{\partial\psi}{\partial y}\frac{dx}{dt})dt=0.\eqno(7)
$$
It also can be written down as follows
$$
\oint\limits_L(\frac{\partial\psi}{\partial x}dy-
\frac{\partial\psi}{\partial y}dx)=0,\eqno(8)
$$
where contour $ L $ is meant as the closed trajectory in space
$ (x, y), $ appropriate to trajectories of considered periodic oscillations of
systems (4). Using the formula of Green [1] expression (8) it is possible to 
transform to a kind:
$$
\int\limits_{\quad S}\int(\frac{\partial^2\psi}{\partial x^2}
+\frac{\partial^2\psi}{\partial y^2})dxdy=0,\eqno(9)
$$
where $ S $ - area on a plane $ (x, y), $ limited by a curve $ L. $
The expressions (7), (8) and (9) represent the various forms of recording 
integrated invariant of system (4).

\vspace*{2mm}

\noindent
{\footnotesize
1.Korn G.A, Korn Th.M. Mathematical handbook for scientists and engineers -- 
Moscow: Nauka, 1974. -- 504p. (in russian) \\
}

\end{document}